\documentclass{article}
\usepackage{amsmath}

\newtheorem{theorem}{Theorem}

\newtheorem{remark}[theorem]{Remark}

\input{tcilatex}

\begin{document}

\begin{center}
{\huge Convenient parameterizations of matrices in terms of vectors}

\bigskip

\emph{M. Bruschi}$^{\ast }$\emph{\ }and\emph{\ F. Calogero}$^{+}$\emph{\ }

Dipartimento di Fisica, Universit\`{a} di Roma "La Sapienza", 00185 Roma,
Italy

Istituto Nazionale di Fisica Nucleare, Sezione di Roma

\bigskip

\textit{Abstract}
\end{center}

Convenient parameterizations of matrices in terms of vectors transform
(certain classes of) matrix equations into covariant (hence
rotation-invariant) vector equations. Certain recently introduced such
parameterizations are tersely reviewed, and new ones introduced.

\bigskip

\bigskip

\bigskip

\bigskip

Mathematics Subject Classification 2000 Index: 15A24; 15A90

Physics and Astronomy Classification Scheme 2001: 02.10.Yn

\bigskip

\bigskip\ 

\bigskip

\bigskip

\bigskip

\bigskip

$^{+}$Corresponding author. Fax: +39-06-4454749

$^{\ast }$mario.bruschi@roma1.infn.it

$^{+}$francesco.calogero@roma1.infn.it, francesco.calogero@uniroma1.it

\newpage

\section{Introduction}

A technique to identify classical (i.e., nonquantal nonrelativistic)
many-body problems amenable to exact treatments in $S$-dimensional space
(with $S>1$) is to firstly identify suitable matrix evolution equations
amenable to exact treatments, and then to parameterize matrices via vectors
so that these matrix evolution equations become \textit{rotation-invariant}
equations of motion of Newtonian type ("acceleration equal force") \cite{1} 
\cite{2} \cite{3}. It is therefore important to identify parameterizations
of matrices in terms of vectors which are suitable to implement this
approach. In this paper we tersely review some representations of this kind
that have been recently used in this context, \cite{4} and we introduce new,
more general and convenient, ones. The exploitation of these latter
representations to identify integrable systems of linear plus cubic
oscillators in $S$-dimensional space (with $S=2,S=3$ as well as arbitrary\ $%
S $) -- analogous yet different from those treated in \cite{2} \cite{3} --
is reported in a separate paper \cite{5}.

\emph{Notation}: hereafter matrices are identified by underlined characters,
and vectors by superimposed arrows; their dimensions should in each case be
clear from the context.

\bigskip

\section{Parameterizations}

Hereafter we denote with the symbol $\ \doteq $ \ the one-to-one
correspondence that the parameterization under consideration institutes
among matrices and $S$-vectors (and, in some cases, scalars). For instance a
well-known parameterization for ($2\otimes 2$)-matrices reads 
\begin{subequations}
\label{1}
\begin{equation}
\underline{M}=\rho \underline{I}+i\overrightarrow{r}\cdot \overrightarrow{%
\underline{\sigma }}  \label{1a}
\end{equation}%
where\ $\rho $\ is a scalar, $\overrightarrow{r}$\ is a $3$-vector, $%
\underline{I}$\ is the unit matrix and the $3$ matrices $\underline{\sigma }%
_{x},\underline{\sigma }_{y}.\underline{\sigma }_{z}$\ are the standard
Pauli matrices,%
\begin{equation}
\underline{I}=%
\begin{pmatrix}
1 & 0 \\ 
0 & 1%
\end{pmatrix}%
,\;\underline{\sigma }_{x}=\left( 
\begin{array}{cc}
0 & 1 \\ 
1 & 0%
\end{array}%
\right) ,\;\underline{\sigma }_{y}=\left( 
\begin{array}{cc}
0 & i \\ 
-i & 0%
\end{array}%
\right) ,\;\underline{\sigma }_{z}=\left( 
\begin{array}{cc}
1 & 0 \\ 
0 & -1%
\end{array}%
\right) ~.  \label{1b}
\end{equation}%
So in this case, in correspondence to (\ref{1a}), we write%
\begin{equation}
\underline{M}\doteq \left( \rho ,\,\overrightarrow{r}\right)  \label{1c}
\end{equation}%
and, via standard calculations, we also have, in self-evident notation,%
\begin{equation}
\underline{M}^{-1}\doteq \frac{\left( \rho ,\overrightarrow{-r}\right) }{%
\rho ^{2}+r^{2}}  \label{1d}
\end{equation}

\begin{equation}
\underline{M}^{\left( 1\right) }\underline{M}^{\left( 2\right) }\doteq
\left( \rho ^{\left( 1\right) }\rho ^{\left( 2\right) }-\overrightarrow{r}%
^{\left( 1\right) }\cdot \overrightarrow{r}^{\left( 2\right) },\,\rho
^{\left( 1\right) }\overrightarrow{r}^{\left( 2\right) }+\rho ^{\left(
2\right) }\overrightarrow{r}^{\left( 1\right) }-\overrightarrow{r}^{\left(
1\right) }\wedge \overrightarrow{r}^{\left( 2\right) }\right)  \label{1e}
\end{equation}

\begin{equation*}
\underline{M}^{\left( 1\right) }\underline{M}^{-1}\underline{M}^{\left(
2\right) }\doteq \left( \rho ^{\left( 1\right) }\rho \rho ^{\left( 2\right)
}+\rho ^{\left( 1\right) }\left( \overrightarrow{r}\cdot \overrightarrow{r}%
^{\left( 2\right) }\right) +\rho ^{\left( 2\right) }\left( \overrightarrow{r}%
\cdot \overrightarrow{r}^{\left( 1\right) }\right) \right.
\end{equation*}%
\begin{equation*}
-\rho \left( \overrightarrow{r}^{\left( 1\right) }\cdot \overrightarrow{r}%
^{\left( 2\right) }\right) +\overrightarrow{r}\cdot \left( \overrightarrow{r}%
^{\left( 1\right) }\wedge \overrightarrow{r}^{\left( 2\right) }\right) ,
\end{equation*}%
\begin{equation*}
\overrightarrow{r}^{\left( 1\right) }\left( \rho \rho ^{\left( 2\right) }+%
\overrightarrow{r}\cdot \overrightarrow{r}^{\left( 2\right) }\right) +%
\overrightarrow{r}^{\left( 2\right) }\left( \rho \rho ^{\left( 1\right) }+%
\overrightarrow{r}\cdot \overrightarrow{r}^{\left( 1\right) }\right)
\end{equation*}%
\begin{equation*}
-\overrightarrow{r}\left( \rho ^{\left( 1\right) }\rho ^{\left( 2\right) }+%
\overrightarrow{r}^{\left( 1\right) }\cdot \overrightarrow{r}^{\left(
2\right) }\right)
\end{equation*}%
\begin{equation}
\left. +\rho ^{\left( 1\right) }\overrightarrow{r}\wedge \overrightarrow{r}%
^{\left( 2\right) }-\rho ^{\left( 2\right) }\overrightarrow{r}\wedge 
\overrightarrow{r}^{\left( 1\right) }-\rho \overrightarrow{r}^{\left(
1\right) }\wedge \overrightarrow{r}^{\left( 2\right) }\right) \,\left( \rho
^{2}+r^{2}\right) ^{-1}~.  \label{1f}
\end{equation}%
And of course introducing $N\otimes N$ block matrices whose elements are ($%
2\otimes 2)-$matrices of type (1a), a parameterization is automatically
introduced of $\left( 4N\right) \otimes (4N)$ matrices in terms of $N^{2}$ $%
3 $-vectors and of $N^{2}$ scalars.

Our approach is analogous but more general: we use (appropriate) block
matrices to introduce (new) parameterizations in terms of $S$-vectors. The
basic block structure of the matrices we parameterize reads as follows: 
\end{subequations}
\begin{equation}
\underline{U}=%
\begin{pmatrix}
\underline{W}^{\left( 11\right) } & \underline{V}^{\left( 11\right) } & 
\underline{W}^{\left( 12\right) } & \underline{V}^{\left( 12\right) } & 
\ldots & \underline{W}^{\left( 1N\right) } & \underline{V}^{\left( 1N\right)
} \\ 
\underline{\tilde{V}}^{\left( 11\right) } & \underline{\tilde{W}}^{\left(
11\right) } & \underline{\tilde{V}}^{\left( 12\right) } & \underline{\tilde{W%
}}^{\left( 12\right) } & \ldots & \underline{\tilde{V}}^{\left( 1N\right) }
& \underline{\tilde{W}}^{\left( 1N\right) } \\ 
\underline{W}^{\left( 21\right) } & \underline{V}^{\left( 21\right) } & 
\underline{W}^{\left( 22\right) } & \underline{V}^{\left( 22\right) } & 
\ldots & \underline{W}^{\left( 2N\right) } & \underline{V}^{\left( 2N\right)
} \\ 
\underline{\tilde{V}}^{\left( 21\right) } & \underline{\tilde{W}}^{\left(
21\right) } & \underline{\tilde{V}}^{\left( 22\right) } & \underline{\tilde{W%
}}^{\left( 22\right) } & \ldots & \underline{\tilde{V}}^{\left( 2N\right) }
& \underline{\tilde{W}}^{\left( 2N\right) } \\ 
\vdots & \vdots & \vdots & \vdots & \ddots & \vdots & \vdots \\ 
\underline{W}^{\left( N1\right) } & \underline{V}^{\left( N1\right) } & 
\underline{W}^{\left( N2\right) } & \underline{V}^{\left( N2\right) } & 
\ldots & \underline{W}^{\left( NN\right) } & \underline{V}^{\left( NN\right)
} \\ 
\underline{\tilde{V}}^{\left( N1\right) } & \underline{\tilde{W}}^{\left(
N1\right) } & \underline{\tilde{V}}^{\left( N2\right) } & \underline{\tilde{W%
}}^{\left( N2\right) } & \ldots & \underline{\tilde{V}}^{\left( NN\right) }
& \underline{\tilde{W}}^{\left( NN\right) }%
\end{pmatrix}
\label{2}
\end{equation}%
The dimensions and structure of the (generally rectangular) matrices $%
\underline{V}^{\left( jk\right) },\underline{\tilde{V}}^{\left( j,k\right) }$%
, $\underline{W}^{\left( jk\right) },\underline{\tilde{W}}^{\left(
j,k\right) }$ $(j,k=1,2,...,N)$ are detailed below, characterizing the
different parameterizations we introduce.

\bigskip

\subsection{Parameterization 1 (P1)}

In this parameterization (see (\ref{2})) the matrices $\underline{V}^{\left(
jk\right) }$ are (generally rectangular) $\left( L\otimes S\right) $%
-matrices, namely matrices with $L$ lines and $S$\ columns, and, conversely,
the matrices $\underline{\tilde{V}}^{\left( j,k\right) }$\ are $\left(
S\otimes L\right) $-matrices, namely matrices with $S$\ lines and $L$\
columns, while the matrices $\underline{W}^{\left( jk\right) },\underline{%
\tilde{W}}^{\left( j,k\right) }$ are identically vanishing square matrices,
more specifically the matrices $\underline{W}^{\left( jk\right) }$ are $%
\left( L\otimes L\right) $-matrices and the matrices $\underline{\tilde{W}}%
^{\left( j,k\right) }$ are $\left( S\otimes S\right) $-matrices. The
consistency of this block structure of the (sparse) $\left( K\otimes
K\right) $-matrix $\underline{U}$, with $K=N\,(L+S)$, is plain.

We obtain a parameterization of the matrix \ $\underline{U}$ in terms of the 
$2\,N^{2}\,L$ $\ S$-vectors $\overrightarrow{r}^{(jk)(\ell )},%
\overrightarrow{\tilde{r}}^{(jk)(\ell )},$%
\begin{subequations}
\label{3}
\begin{equation}
\underline{U}\doteq \left( \overrightarrow{r}^{(jk)(\ell )},~\overrightarrow{%
\tilde{r}}^{(jk)(\ell )}\right) ~,  \label{3a}
\end{equation}%
where of course (here and below, in this section) the indices $j,\,k$ range
from $1$ to $N$ and the index $\ell $ from $1$ to $L$ ($j,k=1,2,..,N;~\ell
=1,2,..,L),$ via the following identifications: 
\begin{equation}
\left( \underline{V}^{\left( jk\right) }\right) _{\ell s}=r_{s}^{(jk)(\ell
)}~,  \label{3b}
\end{equation}

\begin{equation}
\left( \underline{\tilde{V}}^{\left( jk\right) }\right) _{s\ell }=\tilde{r}%
_{s}^{(jk)(\ell )}~.  \label{3c}
\end{equation}%
In the last two formulas the quantities $r_{s}^{(jk)(\ell )}$\ respectively $%
\tilde{r}_{s}^{(jk)(\ell )}$, with the index $s$ ranging of course from $1$
to $S$, are the components of the $N^{2}\,L$ $\ S$-vectors $\overrightarrow{r%
}^{(jk)(\ell )}$ respectively of the $N^{2}\,L$ $\ S$-vectors $%
\overrightarrow{\tilde{r}}^{(jk)(\ell )}$. Now it is straightforward to
verify the following (remarkable) formula:%
\begin{equation}
\underline{U}^{\left[ 1\right] }\underline{U}^{\left[ 2\right] }\underline{U}%
^{\left[ 3\right] }\doteq \left( \overrightarrow{R}^{(jk)(\ell )},%
\overrightarrow{\tilde{R}}^{(jk)(\ell )}\right) ~,  \label{3d}
\end{equation}%
with the $N^{2}\,L$ $\ S$-vectors $\overrightarrow{R}^{(jk)(\ell )},$
respectively the $N^{2}\,L$ $\ S$-vectors $\overrightarrow{\tilde{R}}%
^{(jk)(\ell )},$ defined by the following \textit{covariant }expressions:%
\begin{equation}
\overrightarrow{R}^{(jk)(\ell )}=\sum_{\mu ,\nu =1}^{N}\sum_{\lambda
=1}^{L}\left( \overrightarrow{r}^{\left[ 1\right] (j\mu )(\ell )}\cdot 
\overrightarrow{\tilde{r}}^{\left[ 2\right] (\mu \nu )(\lambda )}\right) \,%
\overrightarrow{r}^{\left[ 3\right] (\nu k)(\lambda )}~,  \label{3e}
\end{equation}%
\begin{equation}
\overrightarrow{\tilde{R}}^{(jk)(\ell )}=\sum_{\mu ,\nu =1}^{N}\sum_{\lambda
=1}^{L}\overrightarrow{\tilde{r}}^{\left[ 1\right] (j\mu )(\lambda
)}\,\left( \overrightarrow{r}^{\left[ 2\right] (\mu \nu )(\lambda )}\cdot 
\overrightarrow{\tilde{r}}^{\left[ 3\right] (\nu k)(\ell )}\right) ~.
\label{3f}
\end{equation}%
Here and throughout a dot sandwiched among two $S$-vectors denotes the
standard scalar product in $S$-dimensional space.

There hold moreover the formulas%
\begin{equation}
\underline{A}\underline{U}\doteq \left( \overrightarrow{R}^{\left( A\right)
(jk)(\ell )},\overrightarrow{\tilde{R}}^{\left( A\right) (jk)(\ell )}\right)
~,  \label{3g}
\end{equation}%
\begin{equation}
\underline{U}\underline{A}\doteq \left( \overrightarrow{R}^{(jk)(\ell
)\left( A\right) },\overrightarrow{\tilde{R}}^{(jk)(\ell )\left( A\right)
}\right)  \label{3h}
\end{equation}%
with%
\begin{equation}
\overrightarrow{R}^{\left( A\right) (jk)(\ell )}=\sum_{\mu =1}^{N}a^{(j\mu
)}\,\overrightarrow{r}^{(\mu k)(\ell )}~~,~~~\overrightarrow{\tilde{R}}%
^{\left( A\right) (jk)(\ell )}=\sum_{\mu =1}^{N}\tilde{a}^{(j\mu )}\,%
\overrightarrow{\tilde{r}}^{(\mu k)(\ell )}~,  \label{3i}
\end{equation}%
\begin{equation}
\overrightarrow{R}^{(jk)(\ell )\left( A\right) }=\sum_{\mu =1}^{N}%
\overrightarrow{r}^{(j\mu )(\ell )}\,\tilde{a}^{(\mu k)}~~,~~~%
\overrightarrow{\tilde{R}}^{(jk)(\ell )\left( A\right) }=\sum_{\mu =1}^{N}%
\overrightarrow{\tilde{r}}^{(j\mu )(\ell )}\,a^{(\mu k)}~,  \label{3j}
\end{equation}%
provided the $\left( K\otimes K\right) $-matrix $\underline{A}$ (with $%
K=N\,(L+S),$ as above) has again the structure (\ref{2}) but now with the $%
N^{2}$ $\ \left( L\otimes S\right) $-matrices $\underline{V}^{\left(
jk\right) },$ as well as the $N^{2}$ $\ \left( S\otimes L\right) $-matrices $%
\underline{\tilde{V}}^{\left( j,k\right) },$ vanishing identically, while
the $N^{2}$ $\ \left( L\otimes L\right) $-matrices $\underline{W}^{\left(
jk\right) }$ are given by%
\begin{equation}
\underline{W}^{\left( jk\right) }=a^{(jk)}\underline{I}  \label{3k}
\end{equation}%
($\underline{I}$ being of course here the $\left( L\otimes L\right) $
identity matrix), and the $N^{2}$ $\left( S\otimes S\right) $-matrices $%
\underline{\tilde{W}}^{\left( jk\right) }$ are given by%
\begin{equation}
\underline{\tilde{W}}^{\left( jk\right) }=\tilde{a}^{(jk)}\underline{I}
\label{3l}
\end{equation}%
($\underline{I}$ being of course here the $\left( S\otimes S\right) $
identity matrix). As suggested by this notation, the $N^{2}$ quantities $%
a^{(jk)},$ as well as the $N^{2}$ quantities $\tilde{a}^{(jk)},$ are
supposed to play the role of \textit{scalars}.

This parameterization was already introduced in \cite{2}; note however that
we use here a somewhat different -- and, we believe, more convenient --
notation.

\bigskip

\subsection{Parameterization 2 (P2)}

In this parameterization (see (\ref{2})) the $N^{2}$ matrices $\underline{V}%
^{\left( jk\right) }$ are $\left( 1\otimes L\right) $-matrices, namely row
matrices, and, conversely, the $N^{2}$ matrices $\underline{\tilde{V}}%
^{\left( j,k\right) }$\ are $\left( L\otimes 1\right) $-matrices, namely
column matrices, while the matrices $\underline{W}^{\left( jk\right) },%
\underline{\tilde{W}}^{\left( j,k\right) }$ are identically vanishing square
matrices, more specifically, the $N^{2}~$(vanishing!) quantities $\underline{%
W}^{\left( jk\right) }\equiv W^{(jk)}$ are $\left( 1\otimes 1\right) $%
-matrices (i. e., \textit{scalars}) while the $N^{2}$ (identically
vanishing!) matrices $\underline{\tilde{W}}^{\left( j,k\right) }$ are $%
\left( L\otimes L\right) $-matrices. The consistency of this block structure
of the (sparse) $\left( K\otimes K\right) $-matrix $\underline{U}$, with $%
K=L\,(N+1)$, is plain ($N,L$ being again \emph{arbitrary} positive
integers). Note that in this parameterization $S=N$. Note that here, and
below, the indices $j,~k,~s$ range from $1$ to $N,$ while the index $\ell $
ranges from $1$ to $L.$

We obtain a parameterization of the matrix \ $\underline{U}$ in terms of the 
$2\,N\,L$ $\ N$-vectors $\overrightarrow{r}^{(n\ell )},\overrightarrow{%
\tilde{r}}^{(n\ell )}$%
\end{subequations}
\begin{subequations}
\label{4}
\begin{equation}
\underline{U}\doteq \left( \overrightarrow{r}^{(n\ell )},\overrightarrow{%
\tilde{r}}^{(n\ell )}\right) ~,  \label{4a}
\end{equation}%
via the following identifications:%
\begin{equation}
\left( \underline{V}^{\left( jn\right) }\right) _{1\ell }=r_{j}^{(n\ell
)}~\;j,n=1,2,..,N;~\ell =1,2,..,L~,  \label{4b}
\end{equation}

\begin{equation}
\left( \underline{\tilde{V}}^{\left( nj\right) }\right) _{\ell 1}=\tilde{r}%
_{j}^{(n\ell )}~\;j,n=1,2,..,N;~\ell =1,2,..,L~.  \label{4c}
\end{equation}%
In the last two formulas the quantities $r_{j}^{(n\ell )}$\ respectively $%
\tilde{r}_{j}^{(n\ell )}$\ are of course the components of the $N$-vectors $%
\overrightarrow{r}^{(n\ell )}$ respectively $\overrightarrow{\tilde{r}}%
^{(n\ell )}$.

Now it is straightforward to verify the following relation:%
\begin{equation}
\underline{U}^{\left[ 1\right] }\underline{U}^{\left[ 2\right] }\underline{U}%
^{\left[ 3\right] }\doteq \left( \overrightarrow{R}^{(n\ell )},%
\overrightarrow{\tilde{R}}^{(n\ell )}\right) ~,  \label{4d}
\end{equation}%
with%
\begin{equation}
\overrightarrow{R}^{(n\ell )}=\sum_{\nu =1}^{N}\sum_{\lambda =1}^{L}%
\overrightarrow{r}^{\left[ 1\right] (\nu \lambda )}\left( \overrightarrow{%
\tilde{r}}^{\left[ 2\right] (\nu \lambda )}\cdot \overrightarrow{r}^{\left[ 3%
\right] (n\ell )}\right) ~,~~~n=1,2,..,N;~\ell =1,2,..,L~,  \label{4e}
\end{equation}%
\begin{equation}
\overrightarrow{\tilde{R}}^{(n\ell )}=\sum_{\nu =1}^{N}\sum_{\lambda
=1}^{L}\left( \overrightarrow{\tilde{r}}^{\left[ 1\right] (n\ell )}\cdot 
\overrightarrow{r}^{\left[ 2\right] (\nu \lambda )}\right) \overrightarrow{%
\tilde{r}}^{\left[ 3\right] (\nu \lambda )}~,~~~n=1,2,..,N;~\ell =1,2,..,L~.
\label{4f}
\end{equation}%
\bigskip The \textit{covariant }structure of these expressions of the $%
2\,N\,L$ \ $N$-vectors $\overrightarrow{R}^{(n\ell )},$ $\overrightarrow{%
\tilde{R}}^{(n\ell )}$ is again remarkable.

And there hold moreover the relations%
\begin{equation}
\underline{A}\underline{U}\doteq \left( \overrightarrow{R}^{\left( A\right)
(n\ell )},\overrightarrow{\tilde{R}}^{\left( A\right) (n\ell )}\right) ~,
\label{4g}
\end{equation}%
\begin{equation}
\underline{U}\underline{A}\doteq \left( \overrightarrow{R}^{(n(\ell )\left(
A\right) },\overrightarrow{\tilde{R}}^{(n\ell )\left( A\right) }\right) ~,
\label{4h}
\end{equation}%
with the $2\,N\,L$ \ $N$-vectors $\overrightarrow{R}^{(A)(n\ell )},$ $%
\overrightarrow{\tilde{R}}^{(A)(n\ell )}$ defined as follows: 
\begin{equation}
\overrightarrow{R}^{\left( A\right) (n\ell )}=\alpha \,\overrightarrow{r}%
^{(n\ell )}\;,\;\overrightarrow{\tilde{R}}^{\left( A\right) (n\ell
)}=\sum_{\nu =1}^{N}\sum_{\lambda =1}^{L}\tilde{a}^{(n,\nu )(\ell \lambda
)}\,\overrightarrow{\tilde{r}}^{(\nu \lambda )}~,~~~n=1,2,..,N;~\ell
=1,2,..,L~,  \label{4i}
\end{equation}%
\begin{equation}
\overrightarrow{R}^{(n\ell )\left( A\right) }=\sum_{\nu =1}^{N}\sum_{\lambda
=1}^{L}\overrightarrow{r}^{(\nu \lambda )}\,\tilde{a}^{(\nu n)(\lambda \ell
)}~~,\;\overrightarrow{\tilde{R}}^{(n\ell )\left( A\right) }=\alpha \,%
\overrightarrow{\tilde{r}}^{(n\ell )}~,~~~n=1,2,..,N;~\ell =1,2,..,L~,
\label{4j}
\end{equation}%
provided the $\left( K\otimes K\right) $-matrix $\underline{A}$ (with $%
K=L\,(N+1))$ has again the structure (\ref{2}) but now with the $\left(
1\otimes L\right) $-matrices $\underline{V}^{\left( jk\right) }$as well as
the $\left( L\otimes 1\right) $-matrices $\underline{\tilde{V}}^{\left(
j,k\right) }$ identically vanishing, while the $\left( 1\otimes 1\right) $%
-matrices $\underline{W}^{\left( jk\right) }$ are given by%
\begin{equation}
\underline{W}^{\left( jk\right) }=\alpha \delta _{jk}~~\;j,k=1,2,..,N~,
\label{4k}
\end{equation}%
($\delta _{jk}$ being of course the Kronecker symbol), and for convenience
we denote as%
\begin{equation}
\tilde{a}^{(jk)(\lambda \ell )}\equiv \left( \underline{\tilde{W}}^{\left(
jk\right) }\right) _{\lambda \ell }~,~~~j,k=1,2,..,N;~\lambda ,\ell =1,2,..,L
\label{4l}
\end{equation}%
the elements of the $\left( L\otimes L\right) $-matrices $\underline{\tilde{W%
}}^{\left( jk\right) }.$

\begin{remark}
Note that with a 'transposed' assignment of the matrices $\underline{V},~%
\underline{\tilde{V}}$ in the above parameterizations (namely, the choice of 
$\left( S\otimes L\right) $-matrices \underline{$V$} and $\left( L\otimes
S\right) $-matrices $\underline{\tilde{V}}$ in the parameterization P1, of
column matrices $\underline{V}$ and row matrices $\underline{\tilde{V}}$ in
the parameterization P2), by changing accordingly the structure of the
matrices $\underline{W},\underline{\tilde{W}},\underline{A}$ one obtains
essentially the same formulas: i.e. the parameterizations are basically the
same, up to a trivial reindexing of the vectors.
\end{remark}

For the readers convenience, we exhibit below the explicit formulas of the
parameterization P2 in the simple case $L=1$ for $2$-vectors, $3$-vectors
and arbitrary $N-$vectors, adding also in each case the important
parameterization for the inverse matrix. To do so, and in order to have a
compact notation for some formulas below, it is convenient to introduce the
external (antisymmetric) product $\overrightarrow{r}^{\left\{ n\right\} }$\
of the $N-1$ \ $N$-vectors obtained excluding the vector $\overrightarrow{r}%
^{\left( n\right) }$ in a set of $N$ \ $N$-vectors $\overrightarrow{r}%
^{\left( k\right) }\equiv \left( r_{1}^{\left( k\right) },r_{2}^{\left(
k\right) },...,r_{N}^{\left( k\right) }\right) ,~k=1,2,...,N$ :

\end{subequations}
\begin{subequations}
\label{5}
\begin{equation}
\overrightarrow{r}^{\left\{ n\right\} }=\overrightarrow{r}^{\left( 1\right)
}\wedge \overrightarrow{r}^{\left( 2\right) }\wedge ...\wedge 
\overrightarrow{r}^{\left( n-1\right) }\wedge \overrightarrow{r}^{\left(
n+1\right) }\wedge ...\wedge \overrightarrow{r}^{\left( N-1\right) }\wedge 
\overrightarrow{r}^{\left( N\right) }  \label{5a}
\end{equation}%
\begin{equation}
=%
\begin{vmatrix}
r_{1}^{\left( 1\right) } & r_{2}^{\left( 1\right) } & \ldots & 
r_{N-1}^{\left( 1\right) } & r_{N}^{\left( 1\right) } \\ 
r_{1}^{\left( 2\right) } & r_{2}^{\left( 2\right) } & \ldots & 
r_{N-1}^{\left( 2\right) } & r_{N}^{\left( 2\right) } \\ 
\vdots & \vdots & \ddots & \vdots & \vdots \\ 
r_{1}^{\left( n-1\right) } & r_{2}^{\left( n-1\right) } & \ldots & 
r_{N-1}^{\left( n-1\right) } & r_{N}^{\left( n-1\right) } \\ 
\overrightarrow{e}^{(1)} & \overrightarrow{e}^{(2)} & \ldots & 
\overrightarrow{e}^{(N-1)} & \overrightarrow{e}^{(N)} \\ 
r_{1}^{\left( n+1\right) } & r_{1}^{\left( n+1\right) } & \ldots & 
r_{1}^{\left( n+1\right) } & r_{1}^{\left( n+1\right) } \\ 
\vdots & \vdots & \ddots & \vdots & \vdots \\ 
r_{1}^{\left( N\right) } & r_{2}^{\left( N\right) } & \ldots & 
r_{N-1}^{\left( N\right) } & r_{N}^{\left( N\right) }%
\end{vmatrix}
\label{5b}
\end{equation}%
where the set of $N$ \ $N$-vectors $\{\overrightarrow{e}^{(n)}\}$, $%
n=1,2,...,N$, provides the standard orthonormal basis in the $N$-vectors
space,%
\begin{equation}
\left( \overrightarrow{e}^{(n)}\right) _{j}=\delta _{nj}~.  \label{5c}
\end{equation}%
Of course for $N=3$ the usual vector product for two $3$-vectors is
recovered but note that the above definition is valid also for $N=2$. Also
note that, with this definition, the scalar product $\overrightarrow{r}%
^{\left( n\right) }\cdot \overrightarrow{r}^{\left\{ n\right\} }$ is
independent of the index $n,$ and it coincides with the standard determinant
associated with the set of $N$ \ $N$-vectors $\left\{ \overrightarrow{r}%
^{\left( k\right) }\right\} ,$%
\begin{equation}
\Delta =\overrightarrow{r}^{\left( n\right) }\cdot \overrightarrow{r}%
^{\left\{ n\right\} }=\left\vert 
\begin{array}{ccc}
r_{1}^{(1)} & \cdots & r_{N}^{(1)} \\ 
\vdots & \ddots & \vdots \\ 
r_{1}^{(N)} & \cdots & r_{N}^{(N)}%
\end{array}%
\right\vert ~,  \label{5d}
\end{equation}%
which has of course a well-known geometrical significance.

The relevant formulas in this parameterization read, in self-evident
notation, as follows.

\bigskip

\subsubsection{($4\otimes 4$)-matrices in terms of four $2$-vectors}

\end{subequations}
\begin{subequations}
\label{6}
\begin{equation}
\underline{U}\doteq \left( \overrightarrow{r}^{\left( 1\right) },%
\overrightarrow{r}^{\left( 2\right) };\overrightarrow{\tilde{r}}^{\left(
1\right) },\overrightarrow{\tilde{r}}^{\left( 2\right) }\right) ~,
\label{6a}
\end{equation}%
\begin{equation}
\overrightarrow{r}^{\left( n\right) }\equiv \left( x^{\left( n\right)
},y^{\left( n\right) }\right) ,~\overrightarrow{\tilde{r}}^{\left( n\right)
}\equiv \left( \tilde{x}^{\left( n\right) },\tilde{y}^{\left( n\right)
}\right) ,~~~n=1,\,2~,  \label{6b}
\end{equation}

\begin{equation}
\overrightarrow{r}^{\left\{ 1\right\} }\equiv \left( y^{\left( 2\right)
},-x^{\left( 2\right) }\right) ,~\overrightarrow{r}^{\left\{ 2\right\}
}\equiv \left( -y^{\left( 1\right) },x^{\left( 1\right) }\right) ~
\label{6bb}
\end{equation}%
(with analogous formulas for the tilded vectors),%
\begin{equation}
\Delta =x^{(1)}\,y^{(2)}-x^{(2)}\,y^{(1)},~\tilde{\Delta}=\tilde{x}^{(1)}\,%
\tilde{y}^{(2)}-\tilde{x}^{(2)}\,\tilde{y}^{(1)}~,  \label{bbb}
\end{equation}%
\begin{equation}
\underline{U}=%
\begin{pmatrix}
0 & x^{\left( 1\right) } & 0 & x^{\left( 2\right) } \\ 
\tilde{x}^{\left( 1\right) } & 0 & \tilde{y}^{\left( 1\right) } & 0 \\ 
0 & y^{\left( 1\right) } & 0 & y^{\left( 2\right) } \\ 
\tilde{x}^{\left( 2\right) } & 0 & \tilde{y}^{\left( 2\right) } & 0%
\end{pmatrix}%
~,  \label{6c}
\end{equation}

\begin{equation}
\underline{U}^{-1}\doteq \left( \frac{\overrightarrow{\tilde{r}}^{\left\{
1\right\} }}{\tilde{\Delta}},\,\frac{\overrightarrow{\tilde{r}}^{\left\{
2\right\} }}{\tilde{\Delta}};\,\frac{\overrightarrow{r}^{\left\{ 1\right\} }%
}{\Delta },\,\frac{\overrightarrow{r}^{\left\{ 2\right\} }}{\Delta }\right)
~,  \label{6d}
\end{equation}%
\begin{equation}
\underline{U}^{\left[ 1\right] }\underline{U}^{\left[ 2\right] }\underline{U}%
^{\left[ 3\right] }\doteq \left( \overrightarrow{R}^{(1)},\overrightarrow{R}%
^{(2)};\overrightarrow{\tilde{R}}^{(1)},\overrightarrow{\tilde{R}}%
^{(2)}\right) ~,  \label{6e}
\end{equation}%
\begin{equation}
\overrightarrow{R}^{(1)}=\overrightarrow{r}^{\left[ 1\right] \left( 1\right)
}\left( \overrightarrow{\tilde{r}}^{\left[ 2\right] \left( 1\right) }\cdot 
\overrightarrow{r}^{\left[ 3\right] \left( 1\right) }\right) +%
\overrightarrow{r}^{\left[ 1\right] \left( 2\right) }\left( \overrightarrow{%
\tilde{r}}^{\left[ 2\right] \left( 2\right) }\cdot \overrightarrow{r}^{\left[
3\right] \left( 1\right) }\right) ~,  \label{6f}
\end{equation}%
\begin{equation}
\overrightarrow{R}^{(2)}=\overrightarrow{r}^{\left[ 1\right] \left( 1\right)
}\left( \overrightarrow{\tilde{r}}^{\left[ 2\right] \left( 1\right) }\cdot 
\overrightarrow{r}^{\left[ 3\right] \left( 2\right) }\right) +%
\overrightarrow{r}^{\left[ 1\right] \left( 2\right) }\left( \overrightarrow{%
\tilde{r}}^{\left[ 2\right] \left( 2\right) }\cdot \overrightarrow{r}^{\left[
3\right] \left( 2\right) }\right) ~,  \label{6g}
\end{equation}%
\begin{equation}
\overrightarrow{\tilde{R}}^{(1)}=\left( \overrightarrow{\tilde{r}}^{\left[ 1%
\right] \left( 1\right) }\cdot \overrightarrow{r}^{\left[ 2\right] \left(
1\right) }\right) \overrightarrow{\tilde{r}}^{\left[ 3\right] \left(
1\right) }+\left( \overrightarrow{\tilde{r}}^{\left[ 1\right] \left(
1\right) }\cdot \overrightarrow{r}^{\left[ 2\right] \left( 2\right) }\right) 
\overrightarrow{\tilde{r}}^{\left[ 3\right] \left( 2\right) }~,  \label{6h}
\end{equation}

\begin{equation}
\overrightarrow{\tilde{R}}^{(2)}=\left( \overrightarrow{\tilde{r}}^{\left[ 1%
\right] \left( 2\right) }\cdot \overrightarrow{r}^{\left[ 2\right] \left(
1\right) }\right) \overrightarrow{\tilde{r}}^{\left[ 3\right] \left(
1\right) }+\left( \overrightarrow{\tilde{r}}^{\left[ 1\right] \left(
2\right) }\cdot \overrightarrow{r}^{\left[ 2\right] \left( 2\right) }\right) 
\overrightarrow{\tilde{r}}^{\left[ 3\right] \left( 2\right) }~,  \label{6i}
\end{equation}%
\begin{equation}
\underline{U}\underline{A}\doteq \left( \tilde{\alpha}^{(11)}\overrightarrow{%
r}^{\left( 1\right) }+\tilde{\alpha}^{(21)}\overrightarrow{r}^{\left(
2\right) },\tilde{\alpha}^{(12)}\overrightarrow{r}^{\left( 1\right) }+\tilde{%
\alpha}^{(22)}\overrightarrow{r}^{\left( 2\right) };\alpha \overrightarrow{%
\tilde{r}}^{\left( 1\right) },\alpha \overrightarrow{\tilde{r}}^{\left(
2\right) }\right) ~,  \label{6j}
\end{equation}%
\begin{equation}
\underline{A}\underline{U}\doteq \left( \alpha \overrightarrow{r}^{\left(
1\right) },\alpha \overrightarrow{r}^{\left( 2\right) };\tilde{\alpha}^{(11)}%
\overrightarrow{\tilde{r}}^{\left( 1\right) }+\tilde{\alpha}^{(12)}%
\overrightarrow{\tilde{r}}^{\left( 2\right) },\tilde{\alpha}^{(21)}%
\overrightarrow{\tilde{r}}^{\left( 1\right) }+\tilde{\alpha}^{(22)}%
\overrightarrow{\tilde{r}}^{\left( 2\right) }\right) ~,  \label{6k}
\end{equation}

with%
\begin{equation}
\underline{A}=%
\begin{pmatrix}
\alpha & 0 & 0 & 0 \\ 
0 & \tilde{\alpha}^{(11)} & 0 & \tilde{\alpha}^{(12)} \\ 
0 & 0 & \alpha & 0 \\ 
0 & \tilde{\alpha}^{(21)} & 0 & \tilde{\alpha}^{(22)}%
\end{pmatrix}%
~.  \label{6l}
\end{equation}

\bigskip

\subsubsection{($6\otimes 6$)-matrices in terms of six $3$-vectors}

\end{subequations}
\begin{subequations}
\label{7}
\begin{equation}
\underline{U}\doteq \left( \overrightarrow{r}^{\left( 1\right) },%
\overrightarrow{r}^{\left( 2\right) },\overrightarrow{r}^{\left( 3\right) };%
\overrightarrow{\tilde{r}}^{\left( 1\right) },\overrightarrow{\tilde{r}}%
^{\left( 2\right) },\overrightarrow{\tilde{r}}^{\left( 3\right) }\right) ~,
\label{7a}
\end{equation}%
\begin{equation}
\overrightarrow{r}^{\left( n\right) }\equiv \left( x^{\left( n\right)
},y^{\left( n\right) },z^{\left( n\right) }\right) ,~\overrightarrow{\tilde{r%
}}^{\left( n\right) }\equiv \left( \tilde{x}^{\left( n\right) },\tilde{y}%
^{\left( n\right) },\tilde{z}^{\left( n\right) }\right) ,~~~n=1,\,2,\,3~,
\label{7b}
\end{equation}%
\begin{equation}
\overrightarrow{r}^{\left\{ n\right\} }=\overrightarrow{r}^{\left(
n+1\right) }\wedge \overrightarrow{r}^{\left( n+2\right) },~\overrightarrow{%
\tilde{r}}^{\left\{ n\right\} }=\overrightarrow{\tilde{r}}^{\left(
n+1\right) }\wedge \overrightarrow{\tilde{r}}^{\left( n+2\right)
},~~~n=1,\,2,\,3~~\func{mod}(3)~,  \label{7bb}
\end{equation}%
\begin{equation}
\Delta =\overrightarrow{r}^{\left( 1\right) }\cdot \overrightarrow{r}%
^{\left( 2\right) }\wedge \overrightarrow{r}^{\left( 3\right) },~\tilde{%
\Delta}=\overrightarrow{\tilde{r}}^{\left( 1\right) }\cdot \overrightarrow{%
\tilde{r}}^{\left( 2\right) }\wedge \overrightarrow{\tilde{r}}^{\left(
3\right) }~,  \label{7bbb}
\end{equation}

\begin{equation}
\underline{U}=%
\begin{pmatrix}
0 & x^{\left( 1\right) } & 0 & x^{\left( 2\right) } & 0 & x^{\left( 3\right)
} \\ 
\tilde{x}^{\left( 1\right) } & 0 & \tilde{y}^{\left( 1\right) } & 0 & \tilde{%
z}^{\left( 1\right) } & 0 \\ 
0 & y^{\left( 1\right) } & 0 & y^{\left( 2\right) } & 0 & y^{\left( 3\right)
} \\ 
\tilde{x}^{\left( 2\right) } & 0 & \tilde{y}^{\left( 2\right) } & 0 & \tilde{%
z}^{\left( 2\right) } & 0 \\ 
0 & z^{\left( 1\right) } & 0 & z^{\left( 2\right) } & 0 & z^{\left( 3\right)
} \\ 
\tilde{x}^{\left( 3\right) } & 0 & \tilde{y}^{\left( 3\right) } & 0 & \tilde{%
z}^{\left( 3\right) } & 0%
\end{pmatrix}%
~,  \label{7c}
\end{equation}%
\begin{equation}
\underline{U}^{-1}\doteq \left( \overrightarrow{R}^{(1)},\overrightarrow{R}%
^{(2)},\overrightarrow{R}^{(3)};\overrightarrow{\tilde{R}}^{(1)},%
\overrightarrow{\tilde{R}}^{(2)},\overrightarrow{\tilde{R}}^{(3)}\right) ~,
\label{7d}
\end{equation}%
\begin{equation}
\overrightarrow{R}^{(n)}=\tilde{\Delta}^{-1}\,\overrightarrow{\tilde{r}}%
^{\left( n+1\right) }\wedge \overrightarrow{\tilde{r}}^{\left( n+2\right)
}~,~~~n=1,\,2,\,3~~\func{mod}(3)~,  \label{7e}
\end{equation}

\begin{equation}
\overrightarrow{\tilde{R}}^{(n)}=\Delta ^{-1}\,\overrightarrow{r}^{\left(
n+1\right) }\wedge \overrightarrow{r}^{\left( n+2\right) }~,~~~n=1,\,2,\,3~~%
\func{mod}(3)~,  \label{7f}
\end{equation}%
\begin{equation}
\underline{U}^{\left[ 1\right] }\underline{U}^{\left[ 2\right] }\underline{U}%
^{\left[ 3\right] }\doteq \left( \overrightarrow{R}^{(1)},\overrightarrow{R}%
^{(2)},\overrightarrow{R}^{(3)};\overrightarrow{\tilde{R}}^{(1)},%
\overrightarrow{\tilde{R}}^{(2)},\overrightarrow{\tilde{R}}^{(3)}\right) ~,
\label{7g}
\end{equation}%
\begin{equation}
\overrightarrow{R}^{(n)}=\sum_{k=1}^{3}\overrightarrow{r}^{\left[ 1\right]
(k)}\left( \overrightarrow{\tilde{r}}^{\left[ 2\right] (k)}\cdot 
\overrightarrow{r}^{\left[ 3\right] (n)}\right) ~,~~~n=1,2,3~,  \label{7h}
\end{equation}%
\begin{equation}
\overrightarrow{\tilde{R}}^{(n)}=\sum_{k=1}^{3}\left( \overrightarrow{\tilde{%
r}}^{\left[ 1\right] (n)}\cdot \overrightarrow{r}^{\left[ 2\right]
(k)}\right) \overrightarrow{\tilde{r}}^{\left[ 3\right] (k)}~,~~~n=1,2,3~,
\label{7i}
\end{equation}

\begin{equation}
\underline{A}\underline{U}\doteq \left( \overrightarrow{R}^{\left( A\right)
(1)},\overrightarrow{R}^{\left( A\right) (2)},\overrightarrow{R}^{\left(
A\right) (3)};\overrightarrow{\tilde{R}}^{\left( A\right) (1)},%
\overrightarrow{\tilde{R}}^{\left( A\right) (2)},\overrightarrow{\tilde{R}}%
^{\left( A\right) (3)}\right) ~,  \label{7j}
\end{equation}%
\begin{equation}
\underline{U}\underline{A}\doteq \left( \overrightarrow{R}^{(1)\left(
A\right) },\overrightarrow{R}^{(2)\left( A\right) },\overrightarrow{R}%
^{(3)\left( A\right) };\overrightarrow{\tilde{R}}^{(1)\left( A\right) },%
\overrightarrow{\tilde{R}}^{(2)\left( A\right) },\overrightarrow{\tilde{R}}%
^{(3)\left( A\right) }\right) ~,  \label{7k}
\end{equation}

\begin{equation}
\overrightarrow{R}^{\left( A\right) (n)}=\alpha \overrightarrow{r}^{(n)}~,~~~%
\overrightarrow{\tilde{R}}^{\left( A\right) (n)}=\sum_{\nu =1}^{3}\tilde{%
\alpha}^{(n\nu )}\overrightarrow{\tilde{r}}^{(\nu )}~,~~~n=1,2,3~,
\label{7l}
\end{equation}%
\begin{equation}
\overrightarrow{R}^{(n)\left( A\right) }=\sum_{\nu =1}^{3}\overrightarrow{r}%
^{(\nu )}\tilde{\alpha}^{(\nu n)}~,~~~\overrightarrow{\tilde{R}}^{(n)\left(
A\right) }=\alpha \overrightarrow{\tilde{r}}^{(n)}~,~~~n=1,2,3~,  \label{7m}
\end{equation}

with%
\begin{equation}
\underline{A}=%
\begin{pmatrix}
\alpha & 0 & 0 & 0 & 0 & 0 \\ 
0 & \tilde{\alpha}^{(11)} & 0 & \tilde{\alpha}^{(12)} & 0 & \tilde{\alpha}%
^{(13)} \\ 
0 & 0 & \alpha & 0 & 0 & 0 \\ 
0 & \tilde{\alpha}^{(21)} & 0 & \tilde{\alpha}^{(22)} & 0 & \tilde{\alpha}%
^{(23)} \\ 
0 & 0 & 0 & 0 & \alpha & 0 \\ 
0 & \tilde{\alpha}^{(31)} & 0 & \tilde{\alpha}^{(32)} & 0 & \tilde{\alpha}%
^{(33)}%
\end{pmatrix}%
~.  \label{7n}
\end{equation}

\bigskip

\subsubsection{($2N\otimes 2N$)-matrices in terms of $2N$ $\;N$-vectors}

\end{subequations}
\begin{subequations}
\label{8}
\begin{equation}
\underline{U}\doteq \left( \overrightarrow{r}^{\left( 1\right) },...,%
\overrightarrow{r}^{\left( N\right) };\overrightarrow{\tilde{r}}^{\left(
1\right) },...,\overrightarrow{\tilde{r}}^{\left( N\right) }\right) ~,
\label{8a}
\end{equation}%
\begin{equation}
\overrightarrow{r}^{\left( n\right) }\equiv \left( r_{1}^{\left( n\right)
},r_{2}^{\left( n\right) },...,r_{N}^{\left( n\right) }\right) ,~~~%
\overrightarrow{\tilde{r}}^{\left( n\right) }\equiv \left( \tilde{r}%
_{1}^{\left( n\right) },\tilde{r}_{2}^{\left( n\right) },...,\tilde{r}%
_{N}^{\left( n\right) }\right) ,~~~n=1,...,N~,  \label{8b}
\end{equation}

\begin{equation}
\underline{U}=%
\begin{pmatrix}
0 & r_{1}^{\left( 1\right) } & 0 & r_{1}^{\left( 2\right) } & 0 & \ldots & 0
& r_{1}^{\left( N\right) } \\ 
\tilde{r}_{1}^{\left( 1\right) } & 0 & \tilde{r}_{2}^{\left( 1\right) } & 0
& \tilde{r}_{3}^{\left( 1\right) } & \ldots & \tilde{r}_{N}^{\left( 1\right)
} & 0 \\ 
0 & r_{2}^{\left( 1\right) } & 0 & r_{2}^{\left( 2\right) } & 0 & \ldots & 0
& r_{2}^{\left( N\right) } \\ 
\tilde{r}_{1}^{\left( 2\right) } & 0 & \tilde{r}_{2}^{\left( 2\right) } & 0
& \tilde{r}_{3}^{\left( 2\right) } & \ldots & \tilde{r}_{N}^{\left( 2\right)
} & 0 \\ 
0 & r_{3}^{\left( 1\right) } & 0 & r_{3}^{\left( 2\right) } & 0 & \ldots & 0
& r_{3}^{\left( N\right) } \\ 
\vdots & \vdots & \vdots & \vdots & \vdots & \ddots & \vdots & \vdots \\ 
0 & r_{N}^{\left( 1\right) } & 0 & r_{N}^{\left( 2\right) } & 0 & \ldots & 0
& r_{N}^{\left( N\right) } \\ 
\tilde{r}_{1}^{\left( N\right) } & 0 & \tilde{r}_{2}^{\left( N\right) } & 0
& \tilde{r}_{3}^{\left( N\right) } & \ldots & \tilde{r}_{N}^{\left( N\right)
} & 0%
\end{pmatrix}%
~,  \label{8c}
\end{equation}%
\begin{equation}
\underline{U}^{-1}\doteq \left( \frac{\overrightarrow{\tilde{r}}^{\left\{
1\right\} }}{\tilde{\Delta}},...,\,\frac{\overrightarrow{\tilde{r}}^{\left\{
N\right\} }}{\tilde{\Delta}};\frac{\overrightarrow{r}^{\left\{ 1\right\} }}{%
\Delta },...,\,\frac{\overrightarrow{r}^{\left\{ N\right\} }}{\Delta }\right)
\label{8d}
\end{equation}%
(see (\ref{5}), and of course analogous formulas hold for the tilded
vectors),%
\begin{equation}
\underline{U}^{\left[ 1\right] }\underline{U}^{\left[ 2\right] }\underline{U}%
^{\left[ 3\right] }\doteq \left( \overrightarrow{R}^{(1)},...,%
\overrightarrow{R}^{(N)};\overrightarrow{\tilde{R}}^{(1)},...,%
\overrightarrow{\tilde{R}}^{(N)}\right) ~,  \label{8e}
\end{equation}%
\begin{equation}
\overrightarrow{R}^{(n)}=\sum_{k=1}^{N}\overrightarrow{r}^{\left[ 1\right]
(k)}\left( \overrightarrow{\tilde{r}}^{\left[ 2\right] (k)}\cdot 
\overrightarrow{r}^{\left[ 3\right] (n)}\right) ~,~~~n=1,2,...,N~,
\label{8f}
\end{equation}%
\begin{equation}
\overrightarrow{\tilde{R}}^{(n)}=\sum_{k=1}^{N}\left( \overrightarrow{\tilde{%
r}}^{\left[ 1\right] (n)}\cdot \overrightarrow{r}^{\left[ 2\right]
(k)}\right) \overrightarrow{\tilde{r}}^{\left[ 3\right]
(k)}~,~~~n=1,2,...,N~,  \label{8g}
\end{equation}

\begin{equation}
\underline{A}\underline{U}\doteq \left( \overrightarrow{R}^{\left( A\right)
(1)},...,\overrightarrow{R}^{\left( A\right) (N)};\overrightarrow{\tilde{R}}%
^{\left( A\right) (1)},...,\overrightarrow{\tilde{R}}^{\left( A\right)
(N)}\right) ~,  \label{8h}
\end{equation}%
\begin{equation}
\underline{U}\underline{A}\doteq \left( \overrightarrow{R}^{(1)\left(
A\right) },...,\overrightarrow{R}^{(N)\left( A\right) };\overrightarrow{%
\tilde{R}}^{(1)\left( A\right) },...,\overrightarrow{\tilde{R}}^{(N)\left(
A\right) }\right) ~,  \label{8i}
\end{equation}%
\begin{equation}
\overrightarrow{R}^{\left( A\right) (n)}=\alpha \overrightarrow{r}^{(n)}~,~~~%
\overrightarrow{\tilde{R}}^{\left( A\right) (n)}=\sum_{\nu =1}^{N}\tilde{%
\alpha}^{(n\nu )}\overrightarrow{\tilde{r}}^{(\nu )}~,~~~n=1,2,...,N~,
\label{8j}
\end{equation}%
\begin{equation}
\overrightarrow{R}^{(n)\left( A\right) }=\sum_{\nu =1}^{N}\overrightarrow{r}%
^{(\nu )}\tilde{\alpha}^{(\nu n)}~,~~~\overrightarrow{\tilde{R}}^{(n)\left(
A\right) }=\alpha \overrightarrow{\tilde{r}}^{(n)}~,~~~n=1,2,...,N~,
\label{8k}
\end{equation}

with%
\begin{equation}
\underline{A}=%
\begin{pmatrix}
\alpha & 0 & 0 & 0 & \ldots & 0 & 0 & 0 \\ 
0 & \tilde{\alpha}^{(11)} & 0 & \tilde{\alpha}^{(12)} & \ldots & \tilde{%
\alpha}^{(1~N-1)} & 0 & \tilde{\alpha}^{(1N)} \\ 
0 & 0 & \alpha & 0 & \ldots & 0 & 0 & 0 \\ 
0 & \tilde{\alpha}^{(21)} & 0 & \tilde{\alpha}^{(22)} & \ldots & \tilde{%
\alpha}^{(2~N-1)} & 0 & \tilde{\alpha}^{(2N)} \\ 
\vdots & \vdots & \vdots & \vdots & \ddots & \vdots & \vdots & \vdots \\ 
0 & \tilde{\alpha}^{(N-1~1)} & 0 & \tilde{\alpha}^{(N-1~2)} & \ldots & 
\tilde{\alpha}^{(N-1~N-1)} & 0 & \tilde{\alpha}^{(N-1~N} \\ 
0 & 0 & 0 & 0 & \ldots & 0 & \alpha & 0 \\ 
0 & \tilde{\alpha}^{(N1)} & 0 & \tilde{\alpha}^{(N2)} & \ldots & \tilde{%
\alpha}^{(N~N-1)} & 0 & \tilde{\alpha}^{(NN)}%
\end{pmatrix}%
~.  \label{8l}
\end{equation}

\bigskip

\section{Concluding remarks}

The matrix parameterizations in terms of vectors reported in this paper have
the property to be preserved -- in terms of vectors yielded by \textit{%
covariant} expressions -- if the matrices are multiplied by appropriate
matrices with scalar matrix elements (indicated as $\underline{A}$, see
above), and as well for the product of three matrices, hence, by iteration,
for the product of any odd number of these matrices (possibly interspersed
by matrices of type $\underline{A}$). They are therefore appropriate to
transform matrix equations that only involve such products, into \textit{%
rotation-invariant} vector equations (for examples see \cite{2} \cite{3} 
\cite{5}); of course in such a context it may also be possible, and
interesting, to also consider reductions, characterized by the presence of a
smaller number of vectors than is naturally yielded by these
parameterizations -- because some vectors can be set to zero and/or be
linearly related to each other (provided this is compatible with the time
evolution under consideration).
\end{subequations}


\newpage

\begin{thebibliography}{9}
\bibitem{1} M. Bruschi and F. Calogero, "Solvable and/or integrable and/or
linearizable $N$-body problems in ordinary (three-dimensional) space. I", J.
Nonlinear Math. Phys. \textbf{7} (2000) 303-386.

\bibitem{2} M. Bruschi and F. Calogero, "Integrable systems of quartic
oscillators", Phys. Lett. \textbf{A273} (2000) 173-182.

\bibitem{3} F. Calogero, \textit{Classical many-body problems amenable to
exact treatments}, Lecture Notes in Physics Monograph \textbf{m66},
Springer, 2001.

\bibitem{4} M. Bruschi and F. Calogero, "On the integrability of certain
matrix evolution equations", Phys. Lett. \textbf{A273} (2000) 167-172.

\bibitem{5} M. Bruschi and F. Calogero, "Integrable systems of quartic
oscillators. II", Phys. Lett. \textbf{A} (next paper).
\end{thebibliography}
\end{document}